\documentclass[%
 aps,
 jmp,%
 amsmath,amssymb,
preprint,
]{revtex4-2}
\usepackage{graphicx}
\usepackage{subfigure}
\usepackage{dcolumn}
\usepackage{bm}
\usepackage[mathlines]{lineno}
\usepackage{graphicx}

\usepackage{graphicx}
\usepackage{subfigure}
\usepackage{dcolumn}
\usepackage{bm}
\usepackage[mathlines]{lineno}
\graphicspath{{figure/}}
\usepackage{array}
\usepackage{braket}
\usepackage{diagbox}
\usepackage{amsmath}
\usepackage{appendix}
\usepackage{multirow}
\usepackage{bm,color,bbm}

\usepackage{float}
\newcolumntype{.}{D{.}{.}{-1}}
\usepackage{amsopn}
\usepackage[colorlinks,linkcolor=blue, anchorcolor=blue, urlcolor=blue, citecolor=blue]{hyperref}
\usepackage{epstopdf}

\begin{document}

\title{Simple Quantum Key Distribution using a Stable Transmitter-Receiver Scheme}
\author{Di Ma$^{1}$, Xin Liu$^{1}$, Chunfeng Huang$^{1}$, Huasheng Chen$^{1}$, Huanbin Lin$^{1}$ and Kejin Wei$^{1,2,3,4,*}$}
\address{
$^1$Guangxi Key Laboratory for Relativistic Astrophysics, School of Physical Science and Technology,
Guangxi University, Nanning 530004, China
$^2$Hefei National Laboratory for Physical Sciences at Microscale and Department of Modern Physics, University of Science and Technology of China, Hefei 230026, China
$^3$Shanghai Research Center for Quantum Sciences, Shanghai 201315, China
$^4$Shanghai Branch, CAS Center for Excellence and Synergetic Innovation Center in Quantum Information and Quantum Physics, University of Science and Technology of China, Shanghai 201315, China
}

$^*$Corresponding author: kjwei@gxu.edu.cn

\begin{abstract}
Quantum Key Distribution (QKD) is a technology that allows secure key exchange between two distant users. A widespread adoption of QKD requires the development of simple, low-cost, and stable systems. However, implementation of the current QKD requires a complex self-alignment process during the initial stage and an additional hardware to compensate the environmental disturbances. In this study, we have presented the implementation of a simple QKD with the help of a stable transmitter-receiver scheme, which simplifies the self-alignment and is robust enough to withstand environmental disturbances. In case of the stability test, the implementation system is able to remain stable for 48 hours and exhibits an average quantum bit error rate  of less than 1\% without any feedback control. The scheme is also tested over a fiber spool, obtaining a stable and secure finite key rate of 7.32k bits per second over a fiber spool extending up to 75 km. The demonstrated long-term stability and obtained secure key rate prove that our method of implementation is a promising alternative for practical QKD systems, in particular,  for Cubesat platform and satellite applications. 
\end{abstract}
\maketitle

Quantum key distribution (QKD)~\cite{2020Xu}, as validated by quantum mechanics, is capable of providing everlasting security that does not rely on any future hardware advances. Since its first invention by Bennet and Brassard in 1984~\cite{2014BB84}, QKD has been attracting a lot of interest, and its feasibility has been experimentally verified in optical fiber~\cite{2012Wang,2013Wei,2019Minder,2019WangS}, free space~\cite{2013Nauerth,2013Vallone,2007PhysRevLett.98.010504}, and underwater channels~\cite{2018Bouchard,2019jIN}. The use of field-test QKD networks has been reported in China~\cite{2014Wang}, Japan~\cite{2011Sasaki}  for various applications. More recently, the first quantum science satellite $\emph{Micius}$ was launched, which paved the way for satellite-to-ground QKD~\cite{2017LiaoQKD,2020Yin} and satellite-relayed intercontinental communication~\cite{2018liaoNetwork}.

Recent researches have focused on developing simpler, lower-cost, and more robust QKD implementations for widespread usage.  This has led to the development of self-compensated modulators designed for different photonic degrees of freedom~\cite{2019Costantino,2018Roberts,2018Wang} as well as several simpler methods of implementation of QKD~\cite{2018Gru,2018Boaron}. Moreover, over the years, integrated photonics based on different platforms has also been introduced for stable and miniaturized systems~\cite{2016Ma,2017sibson,2018Buna,2020Wei}. Furthermore, a detailed study of various novel technologies has been done with an aim to: simplify the consumption of temporal synchronization~\cite{2020Calderaro},  directly predict optimal parameters~\cite{2019Wang}, and enhance the speed of active phase-shift or polarization compensation~\cite{2017Ding,2019Liu}.

\begin{figure*}[hbt]
	\centering
	\includegraphics[width=0.8\linewidth]{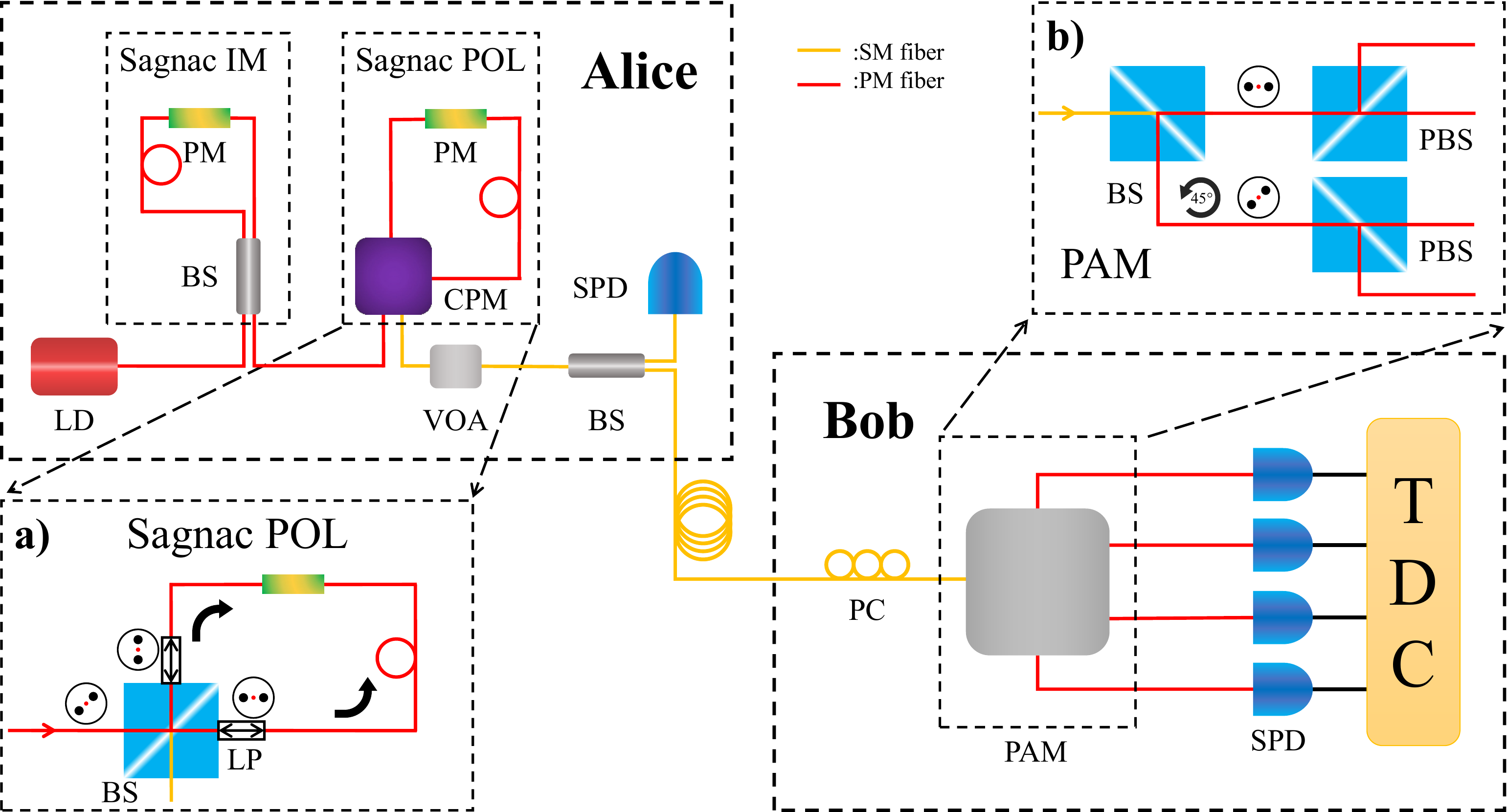}
	\caption{Diagram of experimental setup. LD: Laser Diode; BS: Beam Splitter; PM: Phase Modulator; CPM: Customized Polarization Module; VOA: Variable Optical Attenuator; PC: Polarization Controller; PAM: Polarization Analysis Module;  SPD: Single Photon Detector;  TDC: Time-to Digital Converter; SM fiber: Single Mode fiber; PM fiber: Polarization-Maintaining fiber. \textbf{a)} Diagram of Sagnac POL. LP: Linear Polarizer.\color{black} \textbf{b)} Diagram of PAM. PBS: Polarized Beam Splitter. }
	\label{QKD}
\end{figure*}

Over the years, polarization-encoded QKD has been extensively studied, and it is widely applied in fiber and free space links. A typical polarization-encoded QKD system can be decomposed into two parts: a transmitter (quantum-state preparation, located in Alice) and its receiver (measurement, located in Bob), as shown in Fig.~\ref{QKD}. The transmitter mainly consists of a laser source, an intensity modulator for decoy-state modulation, and a polarization-encoding modulator for quantum-state preparation. The transmitted quantum states are measured by the receiver, which is a conjugated polarization-base analyzer.

In general, for a good performance, a calibration procedure, which usually involves several Polarization Controllers (PCs), is implemented in the transmitter and the receiver while starting the system.  This is based on the fact that although several polarization-encoded modulators (POL)~\cite{2016wangjindong,2014zhiyuan}, particularly POL based on the Sagnac interferometer~\cite{2019Costantino,2019LiYP,2020Agnesi}, exhibit superior stability,
it is still necessary to adjust the PC to balance the intensity of $\left | H  \right \rangle$ and $\left | V  \right \rangle$ components. The stability of the modulator only keep dozens of minutes since the direction of PC is inevitably changed by its mechanical structure and external environment disturbances~\cite{2014zhiyuan,2020Agnesi}. Hence, over time, a re-alignment procedure also occurs constantly.  We also note that several recently proposed Saganc-based structures employing the intensity modulator~\cite{2018Roberts} and POL~\cite{2019Liyang} are free of alignment; however, the application of these individual components in the construction of a QKD system has not been addressed. \color{black} Furthermore, due to the environmental disturbances in the channel, active polarization compensation schemes are typically utilized in a polarization-encoded system. The tracking process takes a certain amount of time since it generally involves two~\cite{2017Ding,2020Agnesi} or more PCs~\cite{2018Lidongdong}.

In this letter, we report a simple method of implementation of QKD using a stable transmitter-receiver scheme.  The transmitter is a combination of the recently proposed Sagnac-intensity~\cite{2018Roberts} and Sagnac-polarization-encoded modulation~\cite{2019Liyang}, which are   connected using a polarization-maintaining pigtail. Hence, the transmitter features free of polarization alignment, and polarization mode dispersion (PDM). Moreover, the transmitter is immune to environmental disturbances. Furthermore, the Sagnac-based intensity modulator possesses the ability to close the security loophole for the recently addressed pattern-effect attack~\cite{2018Yoshino}. In the case of the receiver, the polarization analysis module is compactly designed and inherently aligns the reference frame for each measuring basis.  Hence it is intrinsically stable for measurement bases and simplifies the electrical control of measuring-basis calibration and polarization compensation, because the number of PCs is reduced from two~\cite{2018Lidongdong,2020Agnesi} to one. In the experimental test, our system exhibited 48 hours of stability and an average Quantum Bit Error Rate (QBER) of less than 1\%  without any active compensation or calibration. The system has also been tested over a fiber spool, obtaining a stable and Secure Finite Key Rate (SKR) of 7.32k bits per second (bps) over a fiber spool of 75 km.

Our experimental setup, which is divided into three parts: transmitter (Alice), channel, and receiver (Bob), is illustrated in Fig.~\ref{QKD}. The transmitter comprises a gain-switched laser diode (LD), coupled with a polarization-maintaining pigtail. The LD generates phase-randomized pulses at a repetition rate of 50 MHz  and the pulse width is 200 ps. Subsequently, the light pulses are coupled into a Sagnac-based Intensity Modulator (Sagnac-IM), which is developed using commercial off-the-shelf components, including a polarization-maintaining Beam Splitter (BS) and a Phase Modulator (PM).  The common-path interference mechanism of Saganc-IM ensures that the signal-state modulation is at the peak point of the voltage response curve. Moreover, the desired signal-to-decoy intensity ratio is tuned by tailoring the coupling ratio of the  BS. Compared to the standard commercial lithium niobate intensity modulator, the Sagnac-IM has an inherently stable performance and is immune to pattern-effect attack~\cite{2018Yoshino}. The basic principle of Sagnac-IM has been illustrated in detail in Ref. ~\cite{2018Roberts}.

Subsequently, the light pulses then enter a Sagnac-based polarization modulator (Sagnac-POL). The Sagnac-POL mainly consists of a customized polarized beam splitter module and a PM (see Fig.~\ref{QKD}a) set inside a Sagnac interferometer. The customized module is an in-house design, and the included components are commercially available, although customized assembly technology was adopted. 
Different from the addition of a single mode fiber input in previous Sagnac-based schemes,  the main feature of the proposed Sagnac-based POL is the adoption of a polarizing main fiber input that is directly aligned at $45^\circ$ to the cube BS, followed by two orthogonally aligned linear polarizers. This eliminates the use of the PC and circulator. Thus, the Sagnac-POL not only benefits in facilitating the stability and liberating the PMD from the self-compensating structure, but is also alignment-free and features an improved integration. Further details of the Saganc-POL are  provided in~\cite{2019Liyang}.

The polarization state of the pulse output of the Sagnac-POL can be expressed as:

\begin{equation}
	\left|\psi\right\rangle=\frac{1}{\sqrt{2}}\left(|H\rangle+e^{i(\varphi-\pi)}|V\rangle\right)
\end{equation}
where $\varphi$ is an additional phase set by carefully adjusting the applied voltage on a PM, while  $\pi$ refers to the phase introduced by a BS\color{black}. Eventually, four polarization states can be produced by  fulfilling the requirements of the BB84 protocol on the basis of the key generation ${Z}=\{|0\rangle,|1\rangle\},$ where: $|0\rangle:=\frac{1}{\sqrt{2}}(|H\rangle+|V\rangle),|1\rangle:=\frac{1}{\sqrt{2}}(|H\rangle-|V\rangle) $, and quantum testing basis
$X=\{|+\rangle,|-\rangle\}$, where: $|+\rangle:=\frac{1}{\sqrt{2}}(|H\rangle+i|V\rangle),|-\rangle:=\frac{1}{\sqrt{2}}(|H\rangle-i|V\rangle) $.

Subsequently, the pulses are attenuated to a single photon level with the help of a Variable Optical Attenuator (VOA), and are sent to the receiver located at Bob's stations.

The receiver setup comprises a PC for actively compensating the disturbances induced from the channel, and a customized Polarization Analysis Module (PAM), which has package dimensions of $11\times10\times2$ $cm^3$ and integrates a BS and two  polarization-maintaining Polarized Beam Splitters (PBSs), as shown in Fig.~\ref{QKD}b.
The measure bases for BB84 are inherently aligned with the module, wherein one of the input PMFs of the PBSs is vertically aligned with the BS, while the another one rotates at an angle of  $45^\circ$. The loss of PAM and PC is 1.4 dB. The output of PBSs is detected with the help of four InGaAS single-photon detectors (SPADs, Qasky Co. Ltd., China) with a detection efficiency of about 10\%  , dark count rate of approximately 400 Hz, and gate width of 1 ns. The detections are recorded by the Time-to-Digital Converter (TDC, quTAG100).

\begin{figure}[!hbt]
	\centering
	\includegraphics[width=0.7\linewidth]{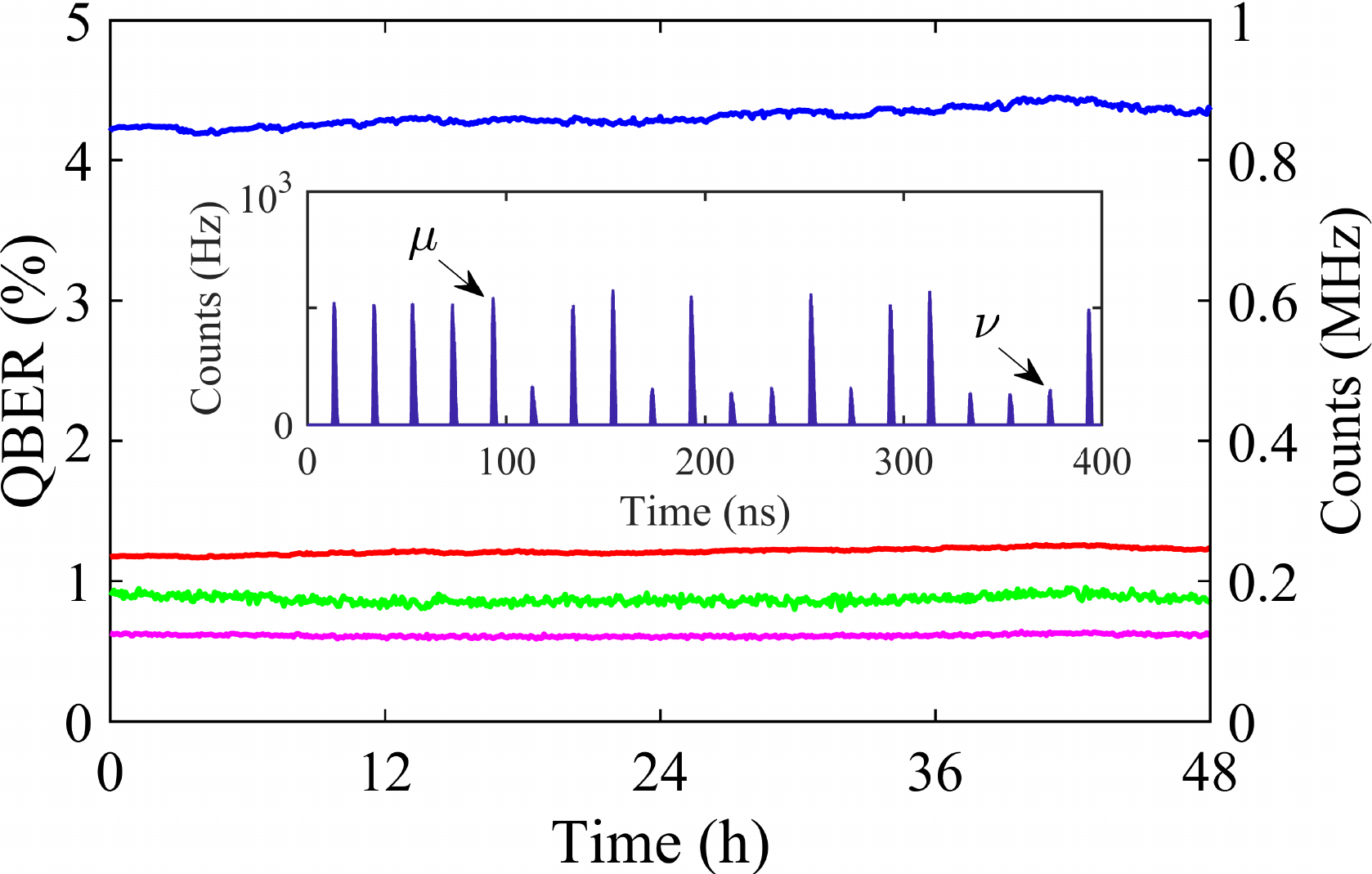}
	\caption{QBER measurement  (left axis) and gains of the signal state and decoy state (right axis) over a 48 hours test.  The green (pink) line denotes QBER of X (Z) basis. The blue (red) lines represents the gains of signal (decoy) state $\mu$ ($\nu$). Data for each basis are collected for 5 seconds at an interval of 5 mins. The inset is the histogram measurement of signal and decoy state using SPADs and TDC, and the y-axis represents the total number of photons located in the time slots. }
	\label{QBER}
\end{figure}

In order to demonstrate the stability of our transmitter and receiver, we measure the overall gain of the signal and weak-decoy state  as well as the QBER for $Z$ and $X$ basis (See Fig.~\ref{QBER}). The measurement is done for random modulations of decoy intensities and polarization qubits  without any re-alignment and feedback control. We perform a one decoy protocol experiment, which was  recently proved in~\cite{2018Rusca}. Accordingly, we choose closed optimum parameters for all distances, wherein the coupling ratio of the BS with respect to the Sagnac-IM was tailored at 25:75 and that related to PAM was set at 90:10. This means that the detection probability of the two measurement bases can be expressed as $P_Z=0.9$ and $P_X=0.1$. To eliminate all the fluctuations not attributable to the study, we removed the fiber spool. Figure~\ref{QBER} clearly reflects the stable gain of each state with an average decoy-signal ratio of $0.2816\pm0.0015$ over a 48 hours test.  Furthermore,  the average QBER measured in case of the X basis is observed to be  $e_X=0.87\pm0.03\%$, \color{black} while that in case of the Z basis was observed to be  $e_Z=0.61\pm0.01\%$. These measurement results clearly demonstrate the high stability of the transmitter and the receiver.

To further test our scheme, we perform a series of one-decoy BB84 QKD  over a commercial fiber spool. For each distance, the secure key rate across a finite key regime is calculated using~\cite{2014Lim,2018Rusca}

\begin{equation}
	\begin{aligned}
		l=&s_{\mathrm{Z}, 0}+s_{\mathrm{Z}, 1}-s_{\mathrm{Z}, 1}h(\phi_{\mathrm{Z}})\\
		&-\lambda_{\mathrm{EC}}-6\log_{2}\frac{19}{\varepsilon_{\mathrm{sec}}}-\log_{2}\frac{2}{\varepsilon_{\mathrm{cor}}},
	\end{aligned}
\end{equation}

where $s_{Z, 0}$ is the lower bound in case of the vacuum events, wherein the Bob registered a detection while the pulse sent by Alice contained no photons. On the other hand, $s_{\mathrm{Z}, 1}$ is the lower bound in case of the single-photon events, defined by the number of detections on the Bob side when the pulse sent by Alice contained only one photon. $\phi_{\mathrm{Z}}$ represents the upper bound on the phase error rate.  $\lambda_{\mathrm{EC}}$ represents the number of disclosed bits in the error correction stage. Meanwhile, $\epsilon_{\mathrm{sec}}$ and $\epsilon_{\mathrm{cor}}$ are the secrecy and correctness parameters, respectively.
\begin{figure}[!hbt]
	\centering
	\includegraphics[width=0.7 \linewidth]{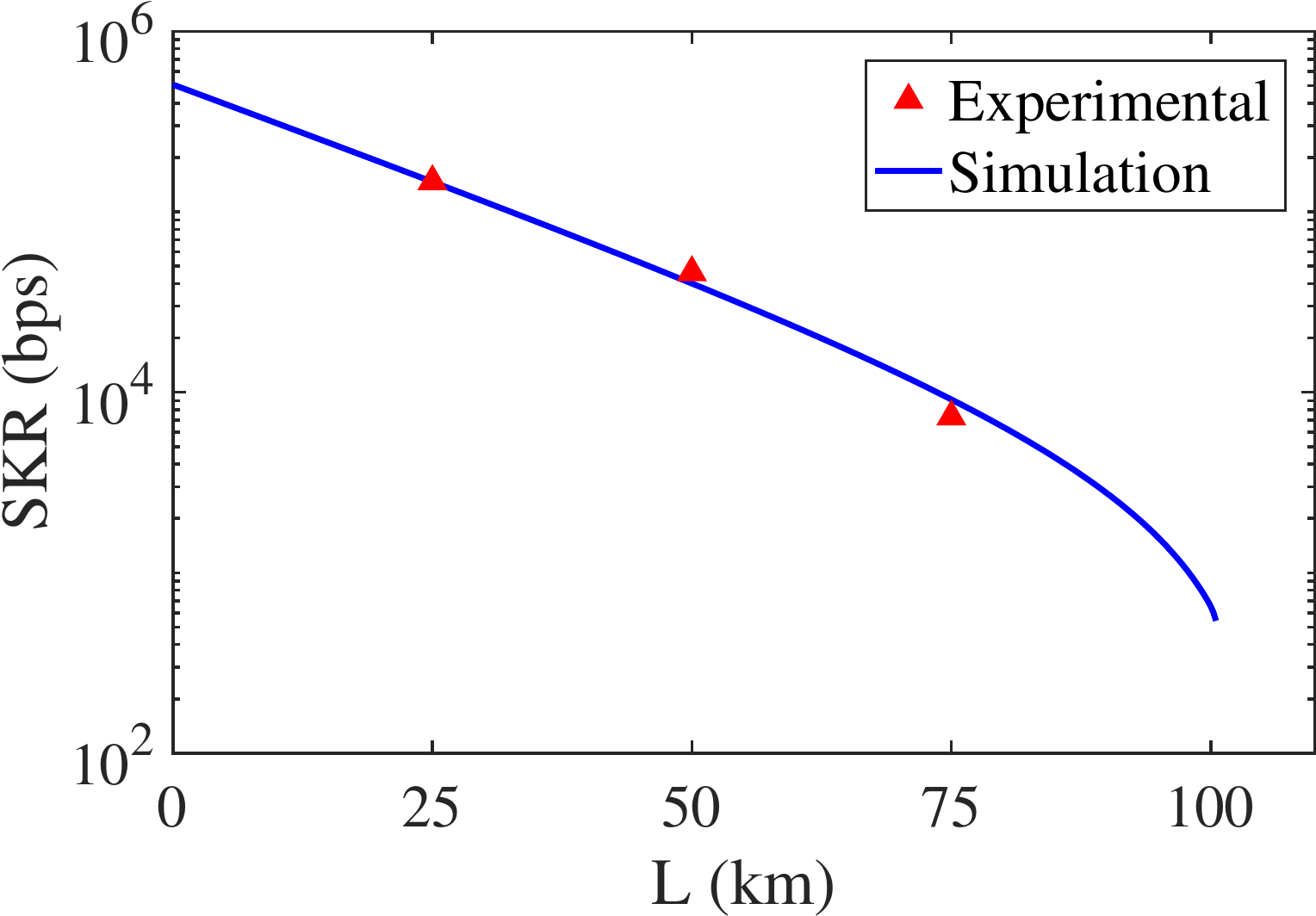}
	\caption{Secure key rate versus fiber distance. The red triangle denotes the experimental result while the blue line denotes the theoretical simulation result based on our experimental parameters.   The channel loss at 25 km, 50 km and 75 km is 4.8 dB, 9.6 dB and 14.6 dB, respectively.}
	\label{SKR}
\end{figure}

Across each distance, a total of  $N=10^{10}$ pulses are sent out. We search the optimized $\mu$ and $P_u$ for the concerned distance. For example, at 75 km, the intensities of the signal state and the decoy state are set to $\mu=0.56$ and $\nu=0.14$, respectively. The probability of transmitting the signal state $\mu$ and the decoy state $\nu$ is $P_{\mu}=0.66$ and $P_{\nu}=0.34$, respectively. The probability of choosing the state in $Z~(X)$ basis is 0.9 (0.1).  The used detection probability of the two measurement bases and the ratio of $\mu:\nu$ remain constant in our test, because they were close to the optimal value, in accordance with our simulation. As reported in Fig.~\ref{SKR}, we achieve a SRK of 7.32 kbps over a 75 km fiber spool.

To summarize, we have presented polarization-encoded QKD using a stable transmitter-receiver scheme. As per the study, the observed overall gains and QBERs exhibited long-term stability over a 48 hours test without any active polarization alignment. In the QKD test, a SKR of 7.32 kbps was observed over a 75 km fiber spool. Due to the limited repetition rate and restricted detection efficiency of this experimental QKD system, its achieved key rate still has a certain gap compared to the other state-of-art polarization-encoding QKD systems~\cite{2018Buna,2017Sibsonoptica}, which distill secret keys at a rate estimated in megahertz bps for a given distance of 50 km.

Fortunately, the feasibility and stability of our key devices, the Saganc-IM and Saganc-PM, have been experimentally demonstrated in the form of GHz in Ref.~\cite{2018Roberts} and \cite{2019Liyang}.
Consider that the system operates at a minimum frequency of 1.25 GHz, as  demonstrated in~\cite{2019Liyang,2018Roberts}, while the SPDs are replaced by
SNSPDs, as in Ref.~\cite{2020Agnesi}, with a free-running dark count rate of 500 Hz and a quantum efficiency of 85\%. The optical misalignment and the sent pluses are set to $1\%$ and $10^{10}$, respectively, which are the same values used in our experiment. In this case, we  expect
that the system would be able to obtain a SKR of  11.2 Mbps at a fiber link distance of 50 km and produce a positive SKR for channel losses of up to 40 dB. The performance of the system is comparable with several reported state-of-the-art implementation systems~\cite{2018Buna,2017Sibsonoptica,2020Gru}. At the same time, the system showcases a higher stability.

Our system can be further improved using the recently proposed qubit synchronization technology \emph{Qubit4Sync}~\cite{2020Calderaro} as well as the qubit compensation scheme~\cite{2017Ding,2020Agnesi} without any additional optical components. This would make our implementation more attractive across different operative scenarios, ranging from the urban metropolitan fiber network to free-space QKD links operated via drones~\cite{2020Liu}. Predominantly, our implementation method is a promising alternative for CubeSat platform~\cite{2017OI} and satellite applications~\cite{2017Bedington}, wherein a long-term stability is required. Furthermore, the security of transmitter could be enhanced by performing quantum attack analysis, in particular, addressing a wavelength-dependent BS attack~\cite{2011Li}.

\section{Funding}
by the National Natural Science Foundation of China (No. 61705048, No. 62031024 and No. 11865004) and the Guangxi Science Foundation (Grant No. 2017GXNSFBA198231).

\section{Acknowledgments}
We thank Yang Li, Wei Li, and Feihu Xu for their enlightening discussions.

\section{Disclosures} The authors declare no conflicts of interest.

\bibliography{QKD}
\end{document}